\documentclass[reprint,aps,prd]{revtex4-1}
\usepackage[english]{babel}
\usepackage{amsmath,amssymb,graphicx,bm,lipsum}
\usepackage[colorlinks=true,citecolor=blue,linkcolor=blue,urlcolor=blue]{hyperref} 

\begin{document}

\title{Overlap reduction function for gravitational wave detectors in an expanding Universe}
\author{Qing-Hua Zhu}
\email{zhuqh@itp.ac.cn}
\affiliation{CAS Key Laboratory of Theoretical Physics,
  Institute of Theoretical Physics, Chinese Academy of Sciences,
  Beijing 100190, China}
\affiliation{School of Physical Sciences,
  University of Chinese Academy of Sciences,
  No. 19A Yuquan Road, Beijing 100049, China}

%%%%%%%%%%%%%%%%%%%%%%%%%%%%%%%%%%%% author %%%%%%%%%%%%%%%%%%%%%%%%%%%%%%%%%%%%

\iffalse
  \author{Qing-Guo Huang}
  \email{Corresponding author: huangqg@itp.ac.cn} 
  \affiliation{CAS Key Laboratory of Theoretical Physics,
    Institute of Theoretical Physics, Chinese Academy of Sciences,
    Beijing 100190, China}
  \affiliation{School of Physical Sciences,
    University of Chinese Academy of Sciences,
    No. 19A Yuquan Road, Beijing 100049, China}
  \affiliation{School of Fundamental Physics and Mathematical Sciences
    Hangzhou Institute for Advanced Study, UCAS, Hangzhou 310024, China}
\fi
\begin{abstract}
  Since it was confirmed two decades ago that the expansion of the Universe is accelerating, it would be of theoretical interests to figure out what is the influence from cosmological constant on detection of stochastic gravitational wave background. 
  This paper studies the overlap reduction functions  in de-Sitter space-time for a pair of one-way tracking gravitational wave detectors. It is shown to be non-trivial in an expanding Universe, because the propagation of light along line of sight also has effect on the response of GW detectors. It is found that the expansion of the Universe can enhance the value of magnitude of the overlap reduction functions, when the detector pairs are close to each other. For nanohertz gravitational waves, this effect can dominate the values of overlap reduction functions when the galactic pulsar pairs are separated by milliarcsecond. %It suggests the auto-correlation of GW is much larger than expected, if the expansion of the Universe is considered.
\end{abstract} 

\maketitle 
\section{Introduction}  

It was confirmed two decades ago that the expansion of the Universe is accelerating, and the present Universe is dominated by the dark energy which still remains mysterious \cite{SupernovaSearchTeam:1998fmf,Bernal:2016gxb}. It would be of theoretical interests to figure out what is the influence from the accelerated expansion of the Universe on physical quantities or observables \cite{Tolish:2016ggo,Kehagias:2016zry,Bieri:2017vni,Perlick:2018iye}. 

The first direct detection of gravitational waves (GWs) open up a new window for exploring the Universe \cite{LIGOScientific:2016aoc}. It implies the existence of stochastic gravitational waves background (GWB), which can be originated from inflationary GW \cite{,Grishchuk:1974ny,Starobinsky:1979ty,Caprini:2018mtu}, produced from early-time phase transitions \cite{Witten:1984rs,Hogan:1986qda,NANOGrav:2021flc}, sourced by cosmic string \cite{Vilenkin:1981bx,Hogan:1984is,Vachaspati:1984gt,LIGOScientific:2021nrg}, or formed by superpositions of unresolved individual GW sources such as binary systems \cite{Schneider:2000sg,Farmer:2003pa,Sesana:2008mz,LIGOScientific:2016fpe,LIGOScientific:2017zlf}, 
core-collapse supernovae \cite{10.1093/mnras/283.2.648,Ferrari:1998ut,Buonanno:2004tp,Finkel:2021zgf}, and deformed rotating neutron stars \cite{Owen:1998xg,Ferrari:1998ut}. To date, the experiments for GW detections in a broad frequency band were built or designed \cite{LIGOScientific:2014pky,VIRGO:2014yos,LISA:2017pwj,TianQin:2015yph,KAGRA:2017tir,Kawamura:2020pcg}. In the $10\rm Hz$--$1\rm kHz$ frequency band, the ground-based GW detectors LIGO/Virgo network at current sensitivity did not find evidence of GWB, and therefore presented an upper limits for the GWB \cite{KAGRA:2021kbb,KAGRA:2021mth}. 
In $\rm nHz$ frequency band, the timing pulsar array projects, namely, NANOGrav \cite{Jenet:2009hk}, PPTA \cite{Hobbs:2013aka} and EPTA \cite{2008AIPC..983..633J}, found and confirmed a common spectrum process from the pulsar-timing data sets, and suggested that further evidence for GWB might rely on its angular correlation signature \cite{NANOGrav:2020bcs,Chen:2021rqp,Goncharov:2021oub}. 

With the assumption of isotropic GWB, the cross-correlations of output of GW detectors depend on the angular separation of a pair of detectors, and the angular-dependence is completely described by overlap reduction functions (ORFs) for the pair of the detectors. For GW detector networks made by pulsar timing array, the ORF of GWs is known as Hellings-Downs curve for a pair of pulsars \cite{Hellings:1983fr}. Theoretically, it is necessary to clarify the possible physical causes that can lead to deviations of the Hellings-Downs curve. For instance, it might come from the GWB beyond isotropy approximation \cite{Mingarelli:2013dsa,Himemoto:2019iwd}, polarized GWB \cite{Omiya:2021zif,Omiya:2020fvw,Chu:2021krj}, non-tensor modes from modified gravity \cite{Nishizawa:2009bf,Lee:2010cg,Liang:2021bct,Chen:2021wdo}, non-linear contributions from higher order perturbation theory \cite{Tasinato:2022xyq}. 
Through a careful calculation on the pulsar terms, it also was found that the value for the magnitude of the ORFs can get larger for the pulsar pairs close to each other \cite{Mingarelli:2014xfa,Hu:2022ujx}. It was suggested an additional correlated phase changes between the close pulsar pairs. In the present study, instead of pulsar terms, we will show a similar behavior of the ORFs due to the cosmological constant. 

This paper investigates the ORFs in de-Sitter background for one-way tracking GW detectors, like PTA. 
Here, the GW detectors are set to be co-moving with the expansion of the Universe. In principle, the setup is not correct for PTA, because the motion of a pulsar is dominated by gravitational field in the galaxy. We thus limit the study to theoretical interests, or perhaps, future gravitational wave timing array \cite{Bustamante-Rosell:2021daj}. We present a rigorous formalism for calculating the ORFs, and show that the calculation is not trivial for de-Sitter background, because the propagation of light along line of sight also has effect on the response of GW detectors. It is found that the cosmological constant leads to a much larger value of ORFs for close GW detector pairs compared with the results calculated in Minkowski space-time.

The rest of the paper is organized as follows. In Sec.~\ref{II}, we brief review the dark energy dominated epochs described by the de-Sitter space-time, and present conventions used in the following. In Sec.~\ref{III}, we show how the cosmological perturbations freely propagate in the de-Sitter vacuum. In Sec.~\ref{IV}, we calculate the light ray affected by the GWs by solving perturbed null geodesic equation to the first order in de-Sitter background. In Sec.~\ref{V}, we calculate the ORFs for a pair of one-way tracking GW detectors, and present its deviations from the Hellings-Downs curve. In Sec.~\ref{VI}, the conclusions and discussions are summarized.

\section{Dark energy dominated epochs and the conventions} \label{II}

Due to the accelerating expansion of our Universe, it would be interesting to extend the studies from the previously in Minkowski space-time into the Friedmann-Lemaitre-Robertson-Walker (FLRW) space-time, and figure out the influence from the expansion of the Universe on the observables. In this section, we will brief review the metric for describing the expanding Universe at late time, and present the conventions used in rest of the paper.

In cosmology, the spatially flat FLRW metric is given by
\begin{eqnarray}
  {\rm d} s^2 & = & - {\rm d} t^2 + a^2 (t) ({\rm d} x^2 + {\rm d} y^2 + {\rm d}
  z^2)~,\label{1}
\end{eqnarray}
where the scale factor $a(t)$ describes the evolution of the Universe. The expression of $a(t)$ can be obtained by solving Einstein field equations sourced by perfect fluids, namely,
\begin{eqnarray}
  H^2 & = & \frac{8 \pi G}{3} \rho~, \label{2}
\end{eqnarray}
where $G$ is gravity constant, $H(\equiv{\dot{a}}/a)$ is Hubble parameter, and $\rho$ is matter density in the Universe.  Based on the standard cosmology, the $\Lambda$CDM model, the matters in the Universe consist of relativistic matter $\rho_r$ (radiation or massless particles), non-relativistic matter $\rho_m$ (baryon, or dark matter), and dark energy $\rho_\Lambda$. Thus Eq.~(\ref{2}) can be rewritten in the form of \cite{Weinberg:2008zzc}
\begin{eqnarray}
  H^2 &=& \frac{8 \pi G}{3} (\rho_r+\rho_m+\rho_\Lambda) \nonumber\\  &=:&  H_0^2 \sqrt{\frac{\Omega_r}{a^4} + \frac{\Omega_m}{a^3} +  \Omega_{\Lambda}} ~,\label{3}
\end{eqnarray}
where the $H_0  (\equiv  {8 \pi G \rho_c}/{3})$ is Hubble constant, and the $\rho_c$ is  critical density that describes the average density in the Universe at the present.
In the second equality of Eq.~(\ref{3}), each components $i = r$, $m$ or $\Lambda$ of matter density are given by $\rho_i = (\Omega_i \rho_c)  a^{- 3(1 + w_i)}$, which can be obtained by using the equations of state $\rho = w  p$ and conservation of energy-momentum tensor. Here, the $\Omega_i$ is density fraction, and $\Omega_r+\Omega_m+\Omega_\Lambda=1$. For different compositions, $w_r=1/3$, $w_m=0$, and $w_\Lambda=-1$, which give different equation of state.

Though the constraints from observation on the cosmic microwave background \cite{Planck:2018vyg}, and local measurement of Hubble law in the late-time Universe \cite{SupernovaSearchTeam:1998fmf}, the density fractions $\Omega_r$, $\Omega_m$ and $\Omega_{\Lambda}$ are determined, in which $\Omega_r \ll \Omega_m \lesssim \Omega_{\Lambda}$. For the late time Universe, there exists an epoch $(\Omega_m / \Omega_{\Lambda})^{\frac{1}{3}} < a < 1$, in which the cosmological redshift is lower than 2.3, approximately. It is known as the epoch that the Universe is dominated by the dark energy or the cosmological constant. In this epochs, the scale factor in Eq.~(\ref{2}) takes the form of
\begin{eqnarray}
  a & = & e^{H_0 t}~. \label{4}
\end{eqnarray}
And the metric in Eq.~(\ref{1}) with scalar factor in Eq.~(\ref{4}) is known as de-Sitter space-time. Here, we adopt the convention $t = 0$ for the present time of the Universe, and $t < 0$ for the history of the Universe.

For simplicity, it is more practical to transform the metric into the conformally flat one. Namely, by introducing the conformal time,
\begin{eqnarray}
  \eta & = & \int^t_{- \infty} \frac{{\rm d} t}{a}\nonumber\\
  & = & - \frac{1}{H_0} e^{- H_0 t}~, \label{5}
\end{eqnarray}
the metric in Eq.~(\ref{1}) reduces to
\begin{eqnarray}
  {\rm d} s^2 & = & \frac{1}{(H_0 \eta)^2} (- {\rm d} \eta^2 + {\rm d} x^2 +
  {\rm d} y^2 + {\rm d} z^2)~. \label{6}
\end{eqnarray}
For the present time of the Universe $t = 0$, the conformal time corresponds to $\eta = -
  \frac{1}{H_0}$. And $\eta < - \frac{1}{H_0}$ describes the history of the
Universe. For  $t\rightarrow 0^-$, the Eq.~(\ref{5}) can be expanded in the form of
\begin{equation}
  \eta = -\frac{1}{H_0}+t+\mathcal{O}((H_0 t)^2)~. \label{7}
\end{equation}
In the late time of the Universe, the $\eta+1/H_0$ is equal to the cosmic time $t$, approximately. We will utilize Eq.~(\ref{7})  in the following for identifying the freely-propagating GWs at $t\rightarrow0^-$.

\section{Cosmological perturbations propagation in vacuum}\label{III}

In de-Sitter background, the propagation of the GW is different from that in Minkowski space-time. 
% It might affect the response of GW detectors, if the arms of the GW detectors are far way from each other.
Due to GW detectors constituted by co-moving periodic sources in the Universe, the GWs, in fact, propagate within the GW detector network. Thus, the cosmological constant might affect the response of GW detectors.
In this section, we will show the evolutions of GWs to the first order with the assumption that the metric perturbations are freely-propagating in the de-Sitter background.

The perturbed de-Sitter metric to the first order is given based on helicity decomposition \cite{Weinberg:2008zzc},
\begin{widetext}
  \begin{eqnarray}
    {\rm d} s^2 & = & \frac{1}{(H_0 \eta)^2} \big(- {\rm d} \eta^2 + (\delta_{i      j} (1 - 2 \psi) + 2 \partial_i \partial_j E + \partial_i C_j +    \partial_j C_i + h_{i   j}) {\rm d} x^i {\rm d} x^j\big)~,\label{8}
  \end{eqnarray}
\end{widetext}
where the $\psi$ and  the $E$ are scalar perturbations, the $C_j$ is vector  perturbation providing $\delta^{i j}\partial_j C^i=0$, and the $h_{ij}$ is  tensor perturbation providing $\delta^{i j}\partial_j h_{i j} = \delta^{i j}h_{i j}=0$.  Here we adopt the Synchronous gauge, because the GW detectors are set to be freely-falling in the Universe.

Evaluating the Einstein field equations, we obtain the evolution equations for the first order metric perturbations,
\begin{widetext}
  \begin{subequations}
    \begin{eqnarray}
%      0 & = & \frac{3}{\eta} \psi' - \frac{1}{\eta} \Delta E' + \Delta \psi~,\\
%      0 & = & 2 \partial_j \psi' + \frac{1}{2} \Delta C_j'~,\\
      0 & = & h_{i   j}'' - \frac{2}{\eta} h_{i   j}' - \Delta h_{i
          j} - \frac{2}{\eta} \partial_i C_j' + \partial_i C_j'' -
      \frac{2}{\eta} \partial_j C_i' + \partial_j C_i'' \nonumber\\
      & & + 2 \partial_i \partial_j
      \left( \psi + E'' - \frac{2}{\eta} E' \right)
      + 2 \delta_{i   j} \left( - \frac{4}{\eta} \psi' + 2 \psi'' +
      \frac{2}{\eta} \Delta E' - \Delta E'' - \Delta \psi \right)~.
    \end{eqnarray}
  \end{subequations}
\end{widetext}
By making of using helicity decomposition, we can split the equations in the form of \cite{Chang:2020tji}
\begin{subequations}
  \begin{eqnarray}
%    \frac{3}{\eta} \psi' - \frac{1}{\eta} \Delta E' + \Delta \psi & = & 0~,\\
%    \psi' & = & 0~,\\
%    \Delta C_j' & = & 0~,\\
    h_{i   j}'' - \frac{2}{\eta} h_{i   j}' - \Delta h_{i
        j} & = & 0~,\\
    - \frac{2}{\eta} C_j' + C_j'' & = & 0~,\\
    \psi + E'' - \frac{2}{\eta} E' & = & 0~,\\
    - \frac{4}{\eta} \psi' + 2 \psi'' + \frac{2}{\eta} \Delta E' - \Delta E'' -
    \Delta \psi & = & 0~.
  \end{eqnarray}
\end{subequations}
Simplifying and evaluating above equations in Fourier space, we obtain
\begin{subequations}
  \begin{eqnarray}
    \psi'_{\bm{k}} & = & 0~,\\
    \frac{2}{\eta} E_{\bm{k}}' - E''_{\bm{k}} & = & \psi_{\bm{k}}~,\\
    C_{j, \bm{k}}' & = & 0~,\\
    h_{i   j, \bm{k}}'' - \frac{2}{\eta} h_{i   j,
        \bm{k}}' + k^2 h_{i   j, \bm{k}} & = & 0~.
  \end{eqnarray}\label{11}
\end{subequations}
It shows that the evolution of tensor perturbation $h_{i   j}$ is described by the wave equations in de-Sitter background, while rest of the metric perturbations are not. Since we only consider that the metric perturbations freely propagate  in de-Sitter background, the solutions of above equations can be formally expressed as initial stochastic variables ($\bar{\psi}_{\bm{k}}, \bar{E}_{\bm{k}}, \bar{C}_{j, \bm{k}}, \text{and } \bar{h}_{i j, \bm{k}}$) and temporal transfer functions $T_{\bm{k}, \ast}$, namely,
\begin{subequations}
  \begin{eqnarray}
    \psi_{\bm{k}} & = & T_{\psi, \bm{k}} (\eta)\bar{\psi}_{\bm{k}}~,\\
    E_{\bm{k}} & = & T_{E, \bm{k}} (\eta) \bar{E}_{\bm{k}}~,\\
    C_{j, \bm{k}} & = & T_{C, \bm{k}} (\eta) \bar{C}_{j, \bm{k}}~,\\
    h_{i   j, \bm{k}} & = & T_{h, \bm{k}} (\eta) \bar{h}_{i j, \bm{k}}~. \label{12-4}
  \end{eqnarray}\label{12}
\end{subequations}
The initial stochastic variables contain physical information about how the perturbations are generated and propagates before its reaching the GW detectors. Because of its stochastic nature, the physical information  should be extracted in statistics. The transfer functions describe the propagation of perturbations within the GW detectors, and thus can affect the response of GW detectors.  In Sec.~\ref{IV}, we will explicitly show that the expression of $T_{h,\bm k}$ can affect the response of GW detectors.

By making use of Eqs.~(\ref{11}) and (\ref{12}), we obtain the expressions of the transfer functions $T_{*,\bm k}$ in the form of
\begin{subequations}
  \begin{eqnarray}
    T_{\psi, \bm{k}} & = & T_{\psi,0}~,\\
    T_{E, \bm{k}} & = & T_{\psi,0}\frac{\eta^2}{2}+T_{E,0}~,\label{13-2}\\
    T_{C, \bm{k}} & = & T_{C,0}~,\\
    T_{h, \bm{k}} & = & e^{- i   k \left(\eta+\frac{1}{H_0}\right)} \left(
    \frac{H_0}{k} \left( \frac{H_0}{2 k} + i   \right) (i   k \eta
    + 1) \right. \nonumber\\ & & \left. + \frac{1}{2} \left( \frac{H_0}{k} \right)^2 (i   k \eta - 1)
    e^{2 i k \left( \eta + \frac{1}{H_0}  \right)} \right)~, \label{13-4}
  \end{eqnarray}\label{13}
\end{subequations}
where $T_{\psi,0}$, $T_{E,0}$, and $T_{C,0}$ are integral constants from solving Eqs.~(\ref{11}), and $k\equiv |\bm k|$. In order to obtain an expression of the $T_{h, \bm{k}}$ that could reduce to the results in Minkowski space-time at $t\rightarrow0^-$, we adopt the boundary conditions that $T_{h, \bm{k}} \rightarrow e^{- i  k (\eta+1/H_0)}$ as $\eta \rightarrow - {1}/{H_0}$.  For the transfer function $T_{E, \bm{k}}$ shown in Eq.~(\ref{13-2}), there seems not a physical mechanism for a large amplitude of the $E_{\bm{k}}$ at large $| \eta |$. Therefore, we let $T_{\psi, \bm{k}} = 0$ and $T_{E, \bm{k}} = {\rm{const}}.$.

The transfer functions of scalar and vector perturbations are constant, while  the transfer function of the tensor perturbation $h_{i   j, \bm{k}}$ oscillates with conformal time. The latter one seems to be more interesting, and could interpret the GWB in the Universe. 
Thus, in the following, we would limit our study to tensor perturbations $h_{i j}$.

\section{Propagation of light in the perturbed de-Sitter space-time}\label{IV}

The GWB formulated by the metric perturbations $h_{i j}$ in the space can affect the propagation of light between the GW detectors. Thus, in this section, we will calculate the propagation of light rays in the perturbed de-Sitter space-time. 

Expanding the null geodesic equations to the first order, we have
\begin{eqnarray}
  0 & = & p^{\mu} \nabla_{\mu} p^{\nu}~, \label{14}\\
  0 & = & \delta p^{\mu} \nabla_{\mu} p^{\nu} + p^{\mu} \nabla_{\mu} \delta
  p^{\nu} \nonumber\\
  & & + g^{\nu \rho} \left( \nabla_{\mu} \delta g_{\lambda \rho} -
  \frac{1}{2} \nabla_{\rho} \delta g_{\mu \lambda} \right) p^{\mu} p^{\lambda}~, \label{15}
\end{eqnarray}
where $p^{\mu}$ and $\delta p^{\mu}$ are the background and the first order 4-velocity of the light, respectively, the $g_{\mu \nu}$ and $\delta g_{\mu \nu}$ are the background metric and the first order metric perturbation, respectively, and ${\nabla_\mu}$ is the covariant derivative with respect to the background metric. In Appendix~\ref{B}, the derivation of Eq.~(\ref{15}) is presented.

Using background metric in Eq.~(\ref{6}), the zeroth order geodesic equations in Eqs.~(\ref{14}) can be evaluated to be
\begin{subequations}
  \begin{eqnarray}
    \partial_0 p^0 & = & \frac{2}{\eta} p^0~,\\
    \partial_0 p^i & = & \frac{2}{\eta} p^i~,
  \end{eqnarray}
\end{subequations}
where we have used the normalization condition for the null 4-velocities $p_{\mu} p^{\mu} = 0$. By integration over the conformal time, we obtain the 4-velocities of backward-propagating light rays,
\begin{eqnarray}
  p^{\mu} & = & w_0 \eta^2 (1, - \hat{n}^i)~,\label{17}
\end{eqnarray}
where the normalized vector $\hat{n}_i (\equiv -{p^i}/{p^0})$ is a constant vector, and $w_0$ is an integral constant from the null geodesic equations. In the case of $\eta \rightarrow -1/H_0$, the $w_0/H_0^2$ represents the frequency of a light ray. 
By solving Eq.~(\ref{17}), we obtain the trajectories of the light rays,
\begin{eqnarray}
  x^i (\eta) - x^i_0 = -\hat{n}^i (\eta - \eta_0)~,
\end{eqnarray}
where $(\eta_0, x_0^i)$ represents initial event. 

For the events of distant objects $(\eta_{\rm{emt}}, d \hat{n}^i)$ and the event on the earth $\left( - {1}/{H_0}, 0 \right)$, the trajectories can be formulated as
\begin{eqnarray}
  d & = &  \frac{1}{H_0} + \eta_{\rm{emt}}~,
\end{eqnarray}
where the $d$ is co-moving distance.
Since the redshift of co-moving objects can be given by $1 + z = {1}/{a} = - ({H_0 \eta_{\rm{emt}}})^{-1}$, it is not difficult to find the distance-redshift relation in de-Sitter space-time \cite{Weinberg:2008zzc},
\begin{eqnarray}
  (1 + z) d & = & \frac{z}{H_0}~, \label{20}
\end{eqnarray}
where one can also define the luminosity distance $d_L\equiv(1+z)d$.
%In the following, we prefer using the quantities distance $d$ and redshift $z$, instead of the $\eta$ for observables, to avoid ambiguity from the coordinate dependence.

Using the background 4-velocities $p^\mu$ in Eq.~(\ref{17}), we can further solve the perturbed geodesic equations in Eqs.~(\ref{15}). Since the GW detectors are set to be freely-falling in the Universe, we evaluate the perturbed geodesic in the Synchronous gauge,
\begin{widetext}
  \begin{subequations}
    \begin{eqnarray}
      0 & = & g_{00} p^0 (\partial_0 - \hat{n}^i \partial_i) \delta p^0 +
      \frac{2}{\eta} p_a \delta p^a - \frac{1}{2} (p^0)^2 \hat{n}^a \hat{n}^a
      \partial_0 h_{a   b}~,\label{21-1}\\
      0 & = & p^{\mu} \partial_{\mu} \delta p^j - \frac{2}{\eta} p^0 \delta p^j +
      \left( p^a p^0 g^{j   b} \partial_0 + p^a p^c g^{j   b}
      \partial_c - \frac{1}{2} p^a p^b g^{j   c} \partial_c +
      \frac{2}{\eta} g^{j   b} p^0 p^a \right) h_{a   b}~,\label{21-2}
    \end{eqnarray}\label{21}
  \end{subequations}
\end{widetext}
where the Latin letters denote spatial indices, and we limit our study to the tensor perturbation $h_{i j}$ in above equations.

In order to obtain ORFs of GW detectors, one should solve $\delta p^0 / p^0$ from perturbed geodesic equations.
In Minkowski space-time, the 0-component of Eqs.~(\ref{15}) is enough for the $\delta p^0 / p^0$, namely,
\begin{eqnarray}
  (\partial_0 - \hat{n}^i \partial_i) \left( \frac{\delta p^0}{p^0} \right) &
  = & - \frac{1}{2} \hat{n}^a \hat{n}^a \partial_0 h_{a   b}~. \label{22}
\end{eqnarray}
However, differed from the calculation in Minkowski space-time, the
0-component of the perturbed geodesic equations in Eq.~(\ref{21-1}) can not be solved without knowing the terms $p_a \delta p^a$. It indicates that the ORFs in de-Sitter space-time is non-trivial, because the propagation of light along the direction of $p^a$ also has effect. 

For solving the $\delta p^0/p^0$, we rewrite the Eq.~(\ref{21-2}) by
contracting a vector $p_j$, which leads to
\begin{eqnarray}
  0 & = & \partial_0 \left( \frac{p_j \delta p^j}{(p^0)^2} \right) +
  \frac{2}{\eta} \frac{p_j \delta p^j}{(p^0)^2} - \hat{n}^c \partial_c \left(
  \frac{p_j \delta p^j}{(p^0)^2} \right) \nonumber\\& & + \hat{n}^a \hat{n}^b \left(
  \partial_0 - \frac{1}{2} \hat{n}^j \partial_j + \frac{2}{\eta} \right) h_{a
      b}~, \label{23}
\end{eqnarray}
Expressing Eqs.~(\ref{21-1}) and (\ref{23}) in Fourier space, we obtain
\begin{subequations}
  \begin{eqnarray}
    0 & = & p_0 (\partial_0 - i \hat{n} \cdot k) \delta p^0_{\bm{k}} +
    \frac{2}{\eta} p_a \delta p^a_{\bm{k}} \nonumber\\ & & - \frac{1}{2} (p^0)^2 \hat{n}^a
    \hat{n}^a \partial_0 h_{a   b, \bm{k}}~,\\
    0 & = & \left( \partial_0 + \frac{2}{\eta} - i \hat{n} \cdot k \right)
    \left( \frac{p_j \delta p^j_{\bm{k}}}{(p^0)^2} \right) \nonumber\\& & + \hat{n}^a
    \hat{n}^b \left( \partial_0 - \frac{1}{2} i \hat{n} \cdot k + \frac{2}{\eta}
    \right) h_{a   b, \bm{k}}~,
  \end{eqnarray}  \label{24}
\end{subequations}
where $\delta p_{\bm k}^\mu$ and $\delta h_{a b,\bm k}$ are the Fourier modes of $\delta p^\mu$ and $\delta h_{a b}$, respectively, and $\hat{n}\cdot k \equiv \hat{n}_j k^j$. Because the $p^\mu$ and $\hat{n}_i$ are free of spatial coordinates in de-Sitter background, they thus have no relevance with the $\bm k$ in Fourier space. It should be clarified that the $\hat{n}_i$ represents the directions of the backward-propagating light rays, and the $\hat{k}^i\equiv k^i/{|\bm k|}$ represents the directions of propagation of the gravitational waves $h_{i j}$.
For simplification, we introduce the $\pi_{\bm{k}}$
and $\chi_{\bm{k}}$ in the form of
\begin{subequations}
  \begin{eqnarray}
    \pi_{\bm{k}} & \equiv & \frac{\delta p_{\bm{k}}^0}{\hat{n}^a
    \hat{n}^b \bar{h}_{a   b  , \bm{k}}}~,\label{25-1}\\
    \chi_{\bm{k}} & \equiv & \frac{\frac{1}{(p^0)^2} p_j \delta
    p_{\bm{k}}^j}{\hat{n}^a \hat{n}^b \bar{h}_{a   b  ,
    \bm{k}}}~.
  \end{eqnarray} \label{25}
\end{subequations}
%where $\bar{h}_{a b,\bm k}$ was defined in Eq.~(\ref{12-4}).
Substituting Eqs.~(\ref{12-4}) and (\ref{25}) into
Eqs.~(\ref{24}), we obtain
\begin{subequations}
  \begin{eqnarray}
    \pi_{\bm{k}}' - i (\hat{n} \cdot k) \pi_{\bm{k}} & = & -
    \frac{p^0}{g_{00}} \left( \frac{2}{\eta} \chi_{\bm{k}} - \frac{1}{2}
    T'_{h,\bm{k}} \right)~,\\
    \chi_{\bm{k}}' - i (\hat{n} \cdot k) \chi_{\bm{k}} + \frac{2}{\eta}
    \chi_{\bm{k}} & = & - T'_{h,\bm{k}} + \frac{1}{2} i (\hat{n} \cdot k)
    T_{h,\bm{k}} - \frac{2}{\eta} T_{h,\bm{k}}~.\nonumber\\
    & &
  \end{eqnarray} \label{26}
\end{subequations}
Here, the solution of $\delta p^0/p^0$ depends on the transfer functions $T_{h,\bm k}(\eta)$ within the GW detectors. By making use of Eqs.~(\ref{13-4}) and (\ref{26}), we obtain explicit expression of $\pi_{\bm{k}}/p^0$ in the form of
\begin{widetext}
  \begin{eqnarray}
    \frac{\pi_{\bm k}}{p^0}&=&  -\frac{i }{4 (1+{\hat{n}\cdot\hat{k}})^4}{\left( \frac{H_0}{k} \right)}^3 \left({ \frac{H_0}{k} }+2 i\right)  \left({(\hat{n}\cdot\hat{k})} \left(3 \eta ^3 k^3+8 \eta  k-8 i\right)+\eta ^2 k^2 (\eta  k-i)\right. \nonumber\\
    && \left. +\eta  k {(\hat{n}\cdot\hat{k})}^3 \left(\eta ^2 k^2+2 i \eta  k+2\right)+{(\hat{n}\cdot\hat{k})}^2 \left(3 \eta ^3 k^3+3 i \eta ^2 k^2+10 \eta  k-2 i\right)\right)e^{-\frac{i k}{H_0} \left(H_0\eta + 1 \right) }\nonumber\\
    && \frac{1}{4 ({\hat{n}\cdot\hat{k}}-1)^4}{\left( \frac{H_0}{k} \right)}^4  \left(i {(\hat{n}\cdot\hat{k})} \left(3 \eta ^3 k^3+8 \eta  k+8 i\right)+\eta ^2 k^2 (1-i \eta  k) \right. \nonumber\\
    && \left. +\eta  k {(\hat{n}\cdot\hat{k})}^3 \left(i \eta ^2 k^2+2 \eta  k+2 i\right)   +{(\hat{n}\cdot\hat{k})}^2 \left(-3 i \eta ^3 k^3-3 \eta ^2 k^2-10 i \eta  k+2\right)\right)e^{\frac{i k}{H_0} \left(H_0\eta + 1 \right) }~, \nonumber\\
    &=&  \left(\frac{{(H_0 \eta)}^3}{2 (1+{\hat{n}\cdot\hat{k}})}-\frac{H_0}{k}\left(\frac{i ({H_0 \eta}-4) {(H_0 \eta)}^2 }{4 (1+{\hat{n}\cdot\hat{k}})}+\frac{3 i {(H_0 \eta)}^2 }{2 (1+{\hat{n}\cdot\hat{k}})^2} \right)\right)e^{-\frac{i k}{H_0} \left(H_0\eta + 1 \right) } \nonumber\\
    && - \frac{i {(H_0 \eta)}^3  }{4 (1-{\hat{n}\cdot\hat{k}})}\left( \frac{H_0}{k} \right)e^{\frac{i k}{H_0} \left(H_0\eta + 1 \right) }  + \mathcal{O}\left(\left(\frac{H_0}{k}\right)^2\right)~. \label{27}
  \end{eqnarray}
\end{widetext}
In the second equality, we expand the $\pi_{\bm k}/p^0$ with $H_0 / k \rightarrow 0$ for the leading order effect of the expansion of the Universe. In the zeroth order with $H_0 / k \rightarrow 0$ and $H_0 \eta \rightarrow -1$, the Eq.~(\ref{27}) reduces to
\begin{eqnarray}
  \frac{\pi_{\bm k}}{p^0} \Bigg|_{\tiny \begin{array}{l}
    H_0 / k \rightarrow 0 \\
    H_0 \eta \rightarrow -1
  \end{array}} = - \frac{1}{2 (1 + \hat{n} \cdot \hat{k})} e^{- i   k (\eta + 1/H_0)}~, 
\end{eqnarray}
which is consistent with the results in Minkowski space-time \cite{Hellings:1983fr,Maggiore:2018sht}.
\begin{figure}
  \includegraphics[width=0.95\linewidth]{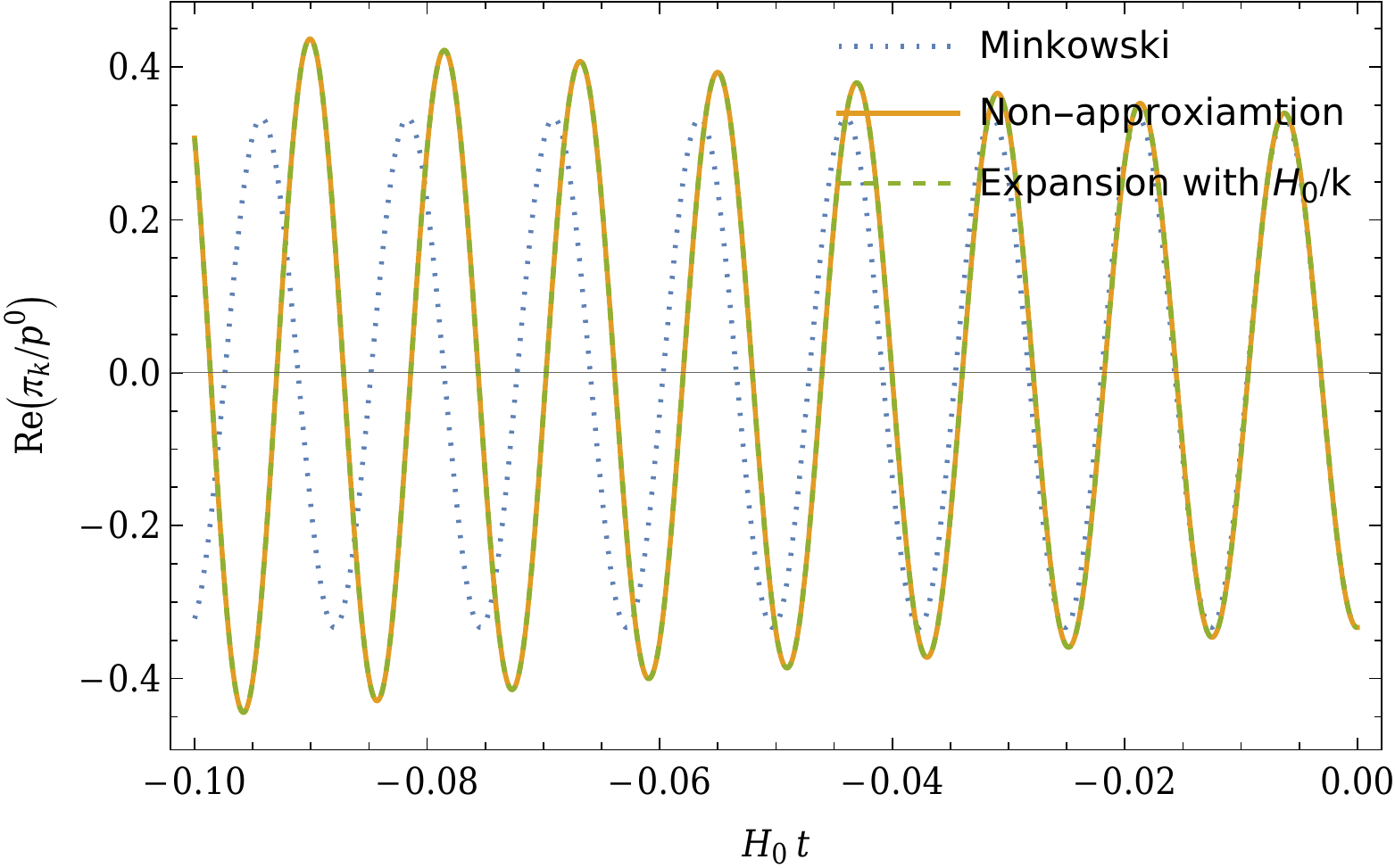}
  \caption{The time evolution of the relative frequency drift of light rays for given frequency $k=H_0/500$. The comparison with the results in Minkowski space-time is presented.}\label{F1}
\end{figure}
In Fig.~\ref{F1}, we show the time evolution of the $\pi_{\bm k}/p^0$ in Eq.~(\ref{27}). Because of expansion of the Universe, there are phase shift and decay of the amplitude for the $\pi_{\bm k}/p^0$.

Finally, by making use of Eqs.~(\ref{25-1}) and (\ref{27}), the frequency drift of the light in configuration space is shown to be
\begin{widetext}
  \begin{eqnarray}
    \frac{\delta p^0}{p^0} & = & \int \frac{{\rm d}^3 k}{(2 \pi)^3} \left\{
    \bar{h}_{a   b, \bm{k}} \hat{n}^a \hat{n}^b
    \frac{\pi_{\bm{k}} (\eta)}{p^0} e^{i   k \cdot x} \right\}\nonumber\\
    & = & \int \frac{{\rm d}^3 k}{(2 \pi)^3} \left\{ \bar{h}_{a   b,
    \bm{k}} \hat{n}^a \hat{n}^b  \left( \left(\frac{{(H_0 \eta)}^3}{2 (1+{\hat{n}\cdot\hat{k}})}-\frac{H_0}{k}\left(\frac{i ({H_0 \eta}-4) {(H_0 \eta)}^2 }{4 (1+{\hat{n}\cdot\hat{k}})}+\frac{3 i {(H_0 \eta)}^2 }{2 (1+{\hat{n}\cdot\hat{k}})^2} \right)\right)e^{-\frac{i k}{H_0} \left(H_0(\eta- \hat{k}\cdot x) + 1 \right) } \right. \right. \nonumber\\
    && \left. \left. - \frac{i {(H_0 \eta) }^3  }{4 (1-{\hat{n}\cdot\hat{k}})}\left( \frac{H_0}{k} \right)e^{\frac{i k}{H_0} \left(H_0(\eta+ \hat{k}\cdot x) + 1 \right) } \right) \right\}~,\nonumber\\
    & & \label{28}
  \end{eqnarray}
\end{widetext}
It is found that the $\delta p^0/p^0$ depends non-linearly on the factor $1/(1+\hat{n}\cdot\hat{k})$, which is even different from the higher order corrections of $\delta p^0/p^0$ \cite{Tasinato:2022xyq}. This might give rise to difficulties in analytical calculations of the improper integral for ORFs.

In the following, we will calculate the ORFs  numerically based on the expression of $\delta p^0/p^0$ shown in above equation.

\section{Gravitational wave detector and overlap function}\label{V}

Providing the isotropic GWB, the ORFs describe the angular correlations of the output of a pair of GW detectors. In this section, we will derive the ORFs for one-way tracking of light with the assumption that the distant clock and receiver both co-moving with the expansion of the Universe. And its deviations from Hellings-Downs curve is also presented.

In de-Sitter background, the time is dilated due to the expansion of the Universe. It can be described by the cosmological redshift,
\begin{eqnarray}
  1 + z & = & \frac{ u_{\mu} p^{\mu} \big|_{\rm{rec}}}{u_{\nu} p^{\nu}  \big|_{\rm{emt}}}~, \label{29}
\end{eqnarray}
where $u_{\mu}$ is the 4-velocities of co-moving objects, and $p^{\mu}$ is the background 4-velocities of light rays. In Synchronous gauge, the 4-velocities of co-moving objects are $u_{\mu} = (- a, 0, 0, 0)$, the subscripts `rec' and `emt' represent the events of receivers, and the event of emitted light from the distant objects, respectively. In this case, the redshift in Eq.~(\ref{29}) reduce to $1+z=p^0_{\rm obs}/(a_{\rm src} p^0_{\rm src})$. For a pulsar as distant clock, its distance from the earth is around $\text{kpc}$. One can estimate its redshift $z \sim 10^{-6}$ based on Eq.~(\ref{20}).

The redshift drift from a distant clock can reflect the space-time fluctuations. Here, it can be derived from the variation of the cosmological redshift,
\begin{eqnarray}
  \Delta z & \equiv & \frac{\delta (1 + z)}{1 + z}\nonumber\\
  & = & \frac{\delta p_{\rm{rec}}^0}{p^0_{\rm{rec}}} - \frac{\delta
  p_{\rm{emt}}^0}{p^0_{\rm{emt}}}~. \label{30}
\end{eqnarray}
In principle, the $\Delta z$ contains the contributions from the perturbed 4-velocities $\delta u_\mu$. In Synchronous gauges, it turns out to be zero.  From Eq.~(\ref{30}), the redshift drift $\Delta z$ depends linearly on the perturbed frequency of light $\delta p^0 / p^0$. %, which we have already obtained in Eq.~(\ref{28}).
The fluctuation of the  distant clock timing can be formulated by the redshift drift of distant objects, because $\Delta t/t=-\Delta f/f$ is independent of specific timing mechanism. Namely, a clock timing by the rotation frequency of a pulsar, or characteristic frequency of an atom must given the same $\Delta t/t$.
We shall clarify that the $\Delta z$ here is redshift drift with respect to the cosmological redshift, while $\Delta z$ was simply called redshift in the calculation of Hellings-Downs curves \cite{Hellings:1983fr,Maggiore:2018sht}. It is because latter one was usually considered in Minkowski space-time, the background redshifts between distant objects are zero, and the leading order redshift comes from the $\Delta z$. They are different physical quantities. The cosmological redshift $z$ can indicate the luminosity distance of co-moving objects, while the redshift drift $\Delta z$ here have no relevance with the distance.

%In practical, one can also define the arrival timing residuals in the form of 
%\begin{eqnarray}
%  R (t) & \equiv & - \int_0^t \Delta z {\rm d} t
%\end{eqnarray}
%where the time $t$ is the cosmic time defined in coordinate $(t, x, y, z)$ shown in Eq.~(\ref{1}). The $t = 0$ represent the current time by convention adopted in Sec.~\ref{III}.

Substituting the expression of $\delta p^0 / p^0$ in Eq.~(\ref{28}) into Eq.~(\ref{30}),
we obtain the redshift drift in the form of
\begin{widetext}
  \begin{eqnarray}
    \Delta z & = &  \int \frac{{\rm d}^3 k}{(2 \pi)^3} \left\{ \bar{h}_{a
    b, \bm{k}} \hat{n}^a \hat{n}^b \left( - \frac{1}{2 (1 + \hat{n} \cdot
        \hat{k})} \left( \mathcal{K}_1 - 3i  \left(\frac{H_0}{k}\right) \mathcal{K}_2 \right) -  \frac{3i}{2(1 + \hat{n}
        \cdot \hat{k})^2} \left( \frac{H_0}{k} \right) \mathcal{K}_3  \right) \right\}~,\label{32}
  \end{eqnarray}
\end{widetext}
where
\begin{subequations}
  \begin{eqnarray}
    \mathcal{K}_1 &\equiv& 1+ (H_0 \eta)^3 e^{-\frac{i k}{H_0}(1+H_0\eta)(1-\hat{n}\cdot\hat{k})}~,\\
    \mathcal{K}_2 &\equiv& 1+ \frac{(H_0 \eta -4)(H_0 \eta)^2}{6} e^{-\frac{i k}{H_0}(1+H_0\eta)(1-\hat{n}\cdot\hat{k})}\nonumber\\
    & & + \frac{(H_0\eta)^3}{6} e^{\frac{i k}{H_0}(1+H_0\eta)(1-\hat{n}\cdot\hat{k})}~,\\
    \mathcal{K}_3 &\equiv& 1- (H_0 \eta)^2 e^{-\frac{i k}{H_0}(1+H_0\eta)(1-\hat{n}\cdot\hat{k})}~.
  \end{eqnarray}
\end{subequations}
%In the case of $\frac{k}{H_0} (1+H_0\eta)> 10$, the oscillation parts of $\Delta z$ is turns out to be 1 as shown in ref.[]. Thus, we simply let these terms to be 1 in the second equal sign of Eq.~(\ref{32}).
Here, the redshifts drift $\Delta z$ is proportional to the tensor perturbation $\bar{h}_{ab,\bm k}$. As shown in Eq.~(\ref{32}), we also limit the calculation to the leading order effects of $H_0/k$.

\begin{figure}[!h]
  \centering
  \includegraphics[scale=0.5]{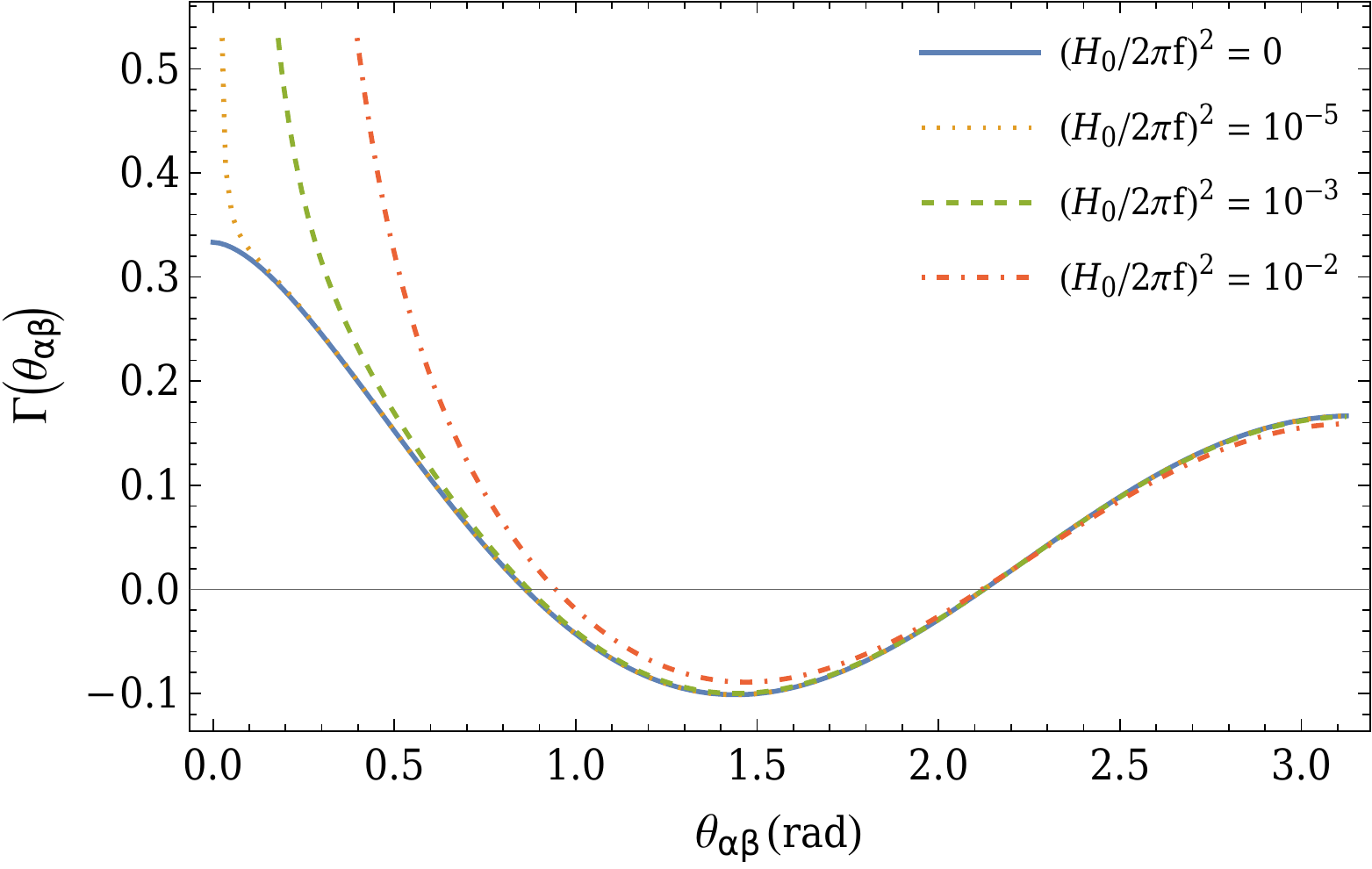}
  \caption{ORFs of GW for different values of $H_0 / (2\pi f)$. The solid curve represents the Hellings-Downs curve.}\label{F2}
\end{figure}

\begin{figure}[!h]
  \centering
  \includegraphics[scale=0.5]{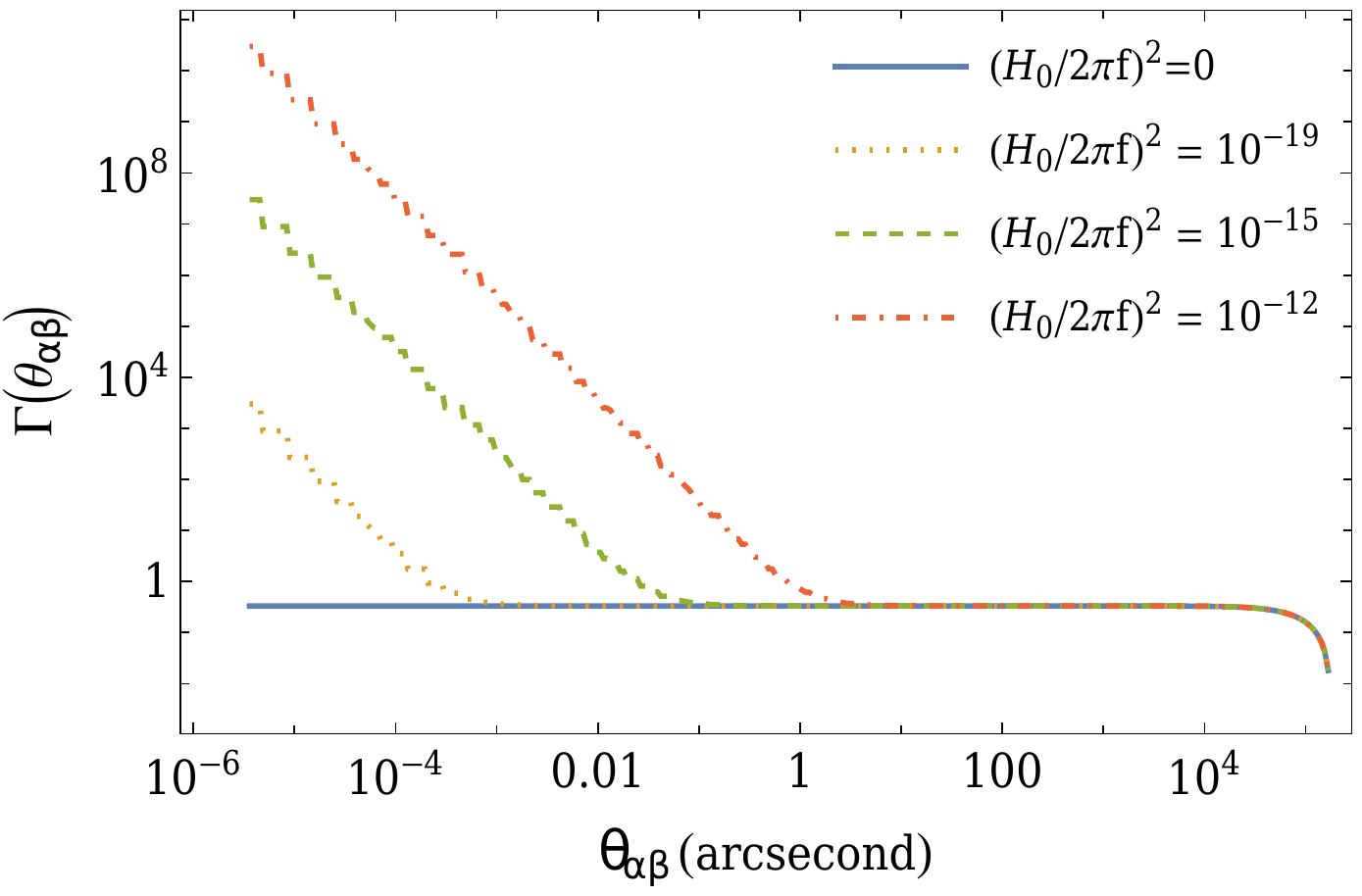}
  \caption{ORFs of GW for different values of $H_0 / (2\pi f)$ in small $\theta_{\alpha,\beta}$.}\label{F3}
\end{figure}

The physical information of stochastic signals of GWBs can be extracted by using cross-correlation functions for the redshift drift from distant clocks $\alpha$ and $\beta$, which are given by
\begin{widetext}
  \begin{eqnarray}
    \langle \Delta z_{\alpha} \Delta z_{\beta} \rangle & = & \int \frac{{\rm d}^3k}{(2 \pi)^3} \int \frac{{\rm d}^3 k'}{(2 \pi)^3} \left\{ \left\langle \bar{h}_{a   b, \bm{k}}^{*} \bar{h}_{c   d,\bm{k}'} \right\rangle \hat{n}^a_{\alpha} \hat{n}^b_{\alpha}\hat{n}^c_{\beta} \hat{n}^c_{\beta} \left( - \frac{1}{2 (1 + \hat{n}_{\alpha} \cdot\hat{k})} \left( \mathcal{K}_{1,\alpha}^* + 3i  \left(\frac{H_0}{k}\right) \mathcal{K}_{2,\alpha}^* \right)
    \right.\right.\nonumber\\ & & \left.\left.
    +  \frac{3i}{2(1 + \hat{n}_{\alpha}\cdot \hat{k})^2} \left( \frac{H_0}{k} \right) \mathcal{K}_{3,\alpha}^* \right)  \left( - \frac{1}{2 (1 + \hat{n}_{\beta} \cdot\hat{k})} \left( \mathcal{K}_{1,\beta} - 3i  \left(\frac{H_0}{k}\right) \mathcal{K}_{2,\beta} \right) -  \frac{3i}{2(1 + \hat{n}_{\beta} \cdot \hat{k})^2} \left( \frac{H_0}{k} \right) \mathcal{K}_{3,\beta}  \right) \right\}\nonumber\\
    & = & \int 4 \pi f^2 {\rm d} f   P_h (f) \int \frac{{\rm d} \Omega}{4
      \pi} \left\{ \Lambda_{a   b, c   d} (\hat{f})
    \hat{n}^a_{\alpha} \hat{n}^b_{\alpha} \hat{n}^c_{\beta} \hat{n}^c_{\beta}
    \right.\nonumber\\ & & \left.
    \times \left( \frac{1}{4 (1 + \hat{n}_{\alpha} \cdot \hat{f}) (1 + \hat{n}_{\beta}\cdot \hat{f})} \left( \mathcal{K}_{1, \alpha}^{ *} \mathcal{K}_{1, \beta}
      +  3i \left(\frac{H_0}{2\pi f}\right) (\mathcal{K}_{2, \alpha}^{ *} \mathcal{K}_{1,
          \beta} -\mathcal{K}_{1, \alpha}^{ *} \mathcal{K}_{2, \beta}) +
      9\left(\frac{H_0}{2\pi f}\right)^2 \mathcal{K}_{2, \alpha}^{ *} \mathcal{K}_{2, \beta}
      \right)
    \right.\right.\nonumber\\ & & \left.\left.
    + \frac{1}{4 (1 + \hat{n}_{\alpha} \cdot \hat{f}) (1 + \hat{n}_{\beta}\cdot \hat{f})^2}\left(\frac{H_0}{2\pi f}\right) \left( 3 i\mathcal{K}_{1,\alpha}^{ *} \mathcal{K}_{3, \beta} - 9\left(\frac{H_0}{2\pi f}\right) \mathcal{K}_{2,\alpha}^{ *} \mathcal{K}_{3, \beta} \right)
    \right.\right.\nonumber\\ & & \left.\left.
    +\frac{1}{4 (1 + \hat{n}_{\alpha} \cdot \hat{f})^2 (1 + \hat{n}_{\beta}\cdot \hat{f})}\left(\frac{H_0}{2\pi f}\right) \left( - 3 i\mathcal{K}_{3, \alpha}^{ *} \mathcal{K}_{1, \beta} -9\left(\frac{H_0}{2\pi f}\right) \mathcal{K}_{3, \alpha}^{ *} \mathcal{K}_{2, \beta}\right)
    \right.\right.\nonumber\\ & & \left.\left.
    \frac{9}{4 (1 + \hat{n}_{\alpha} \cdot \hat{f})^2 (1 + \hat{n}_{\beta}\cdot \hat{f})^2}\left(\frac{H_0}{2\pi f}\right)^2 \mathcal{K}_{3,\alpha}^*\mathcal{K}_{3,\beta}
    \right) \right\}~, \label{35}
  \end{eqnarray}
\end{widetext}
where the ${\rm d} \Omega$ is surface element with respect to $\bm k$, the $\theta_{\alpha \beta}$ is angular distance between the
distant clocks $\alpha$ and $\beta$, the $f$ is present-day physical frequency defined with $f \equiv (2  \pi)^{- 1} k$ and $\hat{f} \equiv \hat{k}$ \cite{LISACosmologyWorkingGroup:2022jok},  the transverse-traceless operator is given by
\begin{eqnarray}
  \Lambda_{a   b, c   d} & \equiv & \mathcal{T}_{a   c}
  \mathcal{T}_{b   d} -\mathcal{T}_{a   b} \mathcal{T}_{c
    d} +\mathcal{T}_{a   d} \mathcal{T}_{b   c}~,
\end{eqnarray}
and above transverse operator $\mathcal{T}_{a   c}$ is defined with
\begin{eqnarray}
  \mathcal{T}_{a   b} (\hat{f}) & = & \delta_{a   b} - \hat{f}_a   \hat{f}_b~.
\end{eqnarray}
Here, we have adopted homogeneous, isotropic and unpolarized GW. And the two-point correlation functions for $\bar{h}_{a b,\bm k}$ can be evaluated to be
\begin{eqnarray}
  \left\langle \bar{h}_{a   b, \bm{k}}^{*} \bar{h}_{c
  d, \bm{k}'} \right\rangle & = & e^{\lambda}_{a   b}
  \left( \bm{k} \right) e^{\lambda'}_{c   d} \left( \bm{k}'
  \right) \left\langle \bar{h}_{\lambda, \bm{k}}^{*}
  \bar{h}_{\lambda', \bm{k}} \right\rangle\nonumber\\
  & = & e^{\lambda}_{a   b} \left( \bm{k} \right) e^{\lambda'}_{c
      d} \left( \bm{k}' \right) (2 \pi)^3 \delta \left( \bm{k} -
  \bm{k}' \right) \delta_{\lambda \lambda'} P_h (k)\nonumber\\
  & = & (2 \pi)^3 \Lambda_{a   b, c   d} (\hat{k}) \delta
  \left( \bm{k} - \bm{k}' \right) P_h (k)~. \label{37}
\end{eqnarray}
%As shown in Eq.~(\ref{34}), the leading order effect from cosmological constant is proportional to the $\left( \frac{H_0}{2\pi f} \right)^2$. %Although the quiet small of the contribution from the cosmological constant, it would be still interesting to explore effect from the expansion of Universe on the angular correlation curved
In the regime $k(1/H_0+ \eta) > 10$, one can take oscillation average that gives $\mathcal{K}^*_{i,\alpha} \mathcal{K}_{i,\beta}\rightarrow 1$. We thus neglect the ``pulsar terms'' from the oscillated parts of $\mathcal{K}_{i,\alpha}$ in Eq.~(\ref{35}).
Therefore, we can read the ORFs in the form of
\begin{widetext}
  \begin{eqnarray}
    \Gamma (\theta_{\alpha \beta}) & = & \int \frac{{\rm d} \Omega}{4 \pi}
    \left\{ \Lambda_{a   b, c   d} (\hat{f}) \hat{n}^a_{\alpha}
    \hat{n}^b_{\alpha} \hat{n}^c_{\beta} \hat{n}^c_{\beta} \left( \frac{1}{4 (1 + \hat{n}_{\alpha} \cdot \hat{f}) (1 + \hat{n}_{\beta}\cdot \hat{f})}
    \right.\right. \nonumber\\ & & \left. \left.
    + \frac{9}{4}\left(\frac{H_0}{2\pi f}\right)^2 \left( \frac{1}{1 + \hat{n}_{\alpha} \cdot \hat{f} }- \frac{1}{(1 + \hat{n}_{\alpha} \cdot \hat{f})^2 } \right)\left( \frac{1}{1 + \hat{n}_{\beta} \cdot \hat{f} }- \frac{1}{(1 + \hat{n}_{\beta} \cdot \hat{f})^2 } \right)
    \right)  \right\}~. \label{39}
  \end{eqnarray}
\end{widetext}
In order to obtain $\Gamma(\theta_{\alpha\beta})$, we can let direction of propagation of gravitational wave as
\begin{eqnarray}
  \hat{f} & = & (\sin \theta \cos \phi, \sin \theta \sin \phi, \cos \theta)~, \label{40}
\end{eqnarray}
where the angular coordinate $\theta$, $\phi$ is defined with respect to $\bm k$, and ${\rm d} \Omega \equiv \sin \theta {\rm d}  \theta {\rm d} \phi$.
Since the angular $\theta_{\alpha \beta}$ is formulated as $\hat{n}_{\alpha}\cdot \hat{n}_{\beta} = \cos \theta_{\alpha \beta}$, the locations of the distant clocks $\alpha$, and $\beta$ can be
\begin{subequations}
\begin{eqnarray}
  \hat{n}_{\alpha} & = & (0, 0, 1)~,\\
  \hat{n}_{\beta} & = & (\sin\theta_{\alpha \beta}, 0, \cos \theta_{\alpha\beta})~.
\end{eqnarray}
\end{subequations}
From Eq.~(\ref{39}), the leading order effect from the expansion of the Universe is proportional to $(H_0/f)^2$ for the ORFs.
By making use of the expression of the $\hat{f}$ in Eq.~(\ref{40}), the transverse-traceless operator acting on $\hat{n}_{\alpha}$ and $\hat{n}_{\beta}$ in Eq.~(\ref{39}) can be evaluated,
\begin{eqnarray}
  \Lambda_{a   b, c   d} \hat{n}^a_{\alpha}  \hat{n}^b_{\alpha} \hat{n}^c_{\beta} \hat{n}_{\beta}^d & = & \frac{1}{4}  \sin^2 \theta \Big((3 + \cos (2 \theta)) \cos (2 \phi) \sin^2 \theta_{\alpha    \beta} \nonumber\\&&+ \sin \theta \big(\sin \theta + 3 \cos (2 \theta_{\alpha \beta}) \sin  \theta \nonumber\\&&- 4 \cos \theta \cos \phi \sin (2 \theta_{\alpha \beta})\big)\Big)~,
\end{eqnarray}
and 
\begin{subequations}
\begin{eqnarray}
  \hat{n}_{\alpha} \cdot \hat{f} & = & \cos \theta~,\\
  \hat{n}_{\beta} \cdot \hat{f} & = & \cos \theta \cos \phi \sin  \theta_{\alpha \beta} + \cos \theta \cos \theta_{\alpha \beta}~.
\end{eqnarray}
\end{subequations}
%\begin{eqnarray}
%  \Gamma (\theta_{\alpha \beta}) & = &
%\end{eqnarray}
In Fig.~{\ref{F2}}, we present the ORFs for different values of $H_0 / (2\pi f)$. It shows that the cosmological constant could enhance the value of the magnitude of ORFs in the case of $\theta_{\alpha \beta} \rightarrow 0$. Similar behavior was also found from a careful calculation on the pulsar terms \cite{Mingarelli:2014xfa,Hu:2022ujx}. In Fig.~{\ref{F3}}, we zoom in the angular correlation curves for small angle $\theta_{\alpha\beta}$. For $n\text{Hz}$ GWB in the PTA band, the $(H_0/k)^2$ is estimated to be $10^{-19}$. In this case, the enhanced values of the ORFs are shown to be dominated for the pulsars pairs that are separated by angular distance less than $m$as. This conclusions can be numerically presented by $\Gamma\left(\theta_{\alpha\beta} \rightarrow 0 \right) \propto \theta_{\alpha \beta}^{-2}$. 

\

\

\section{Conclusions and discussions}\label{VI}

We investigated the ORFs in de-Sitter background for one-way tracking GW detectors. 
It was found that the cosmological constant leads to a much larger value of magnitude of the ORFs, when GW detector pairs are close to each other. For nanohertz gravitational waves, this effect can dominate of value of ORFs when the galactic pulsar pairs are separated by milliarcsecond. We calculate the ORFs in de-Sitter background for the first time. For GW detections in low frequency band in the future, it is inevitable to be confront with the effect from the expansion of the Universe, such as the gravitational wave timing array constituted by distant binaries \cite{Bustamante-Rosell:2021daj}.

From the comparison between Eqs.~(\ref{21}) and Eq.~(\ref{22}), calculation on the ORFs in de-Sitter background is shown to be non-trivial, because spatial components of perturbed 4-velocities of light also have effect on the frequency drift for GW detectors. Thus, one can not account for the difference of the ORFs between those in an expansion Universe and Minkowski space-time from simply a redshift factor $1+z$. 

The GW detectors are set to be co-moving with the expansion of the Universe, which is not suited for describing PTA, because the motion of a pulsar is dominated by gravitational field in the galaxy. Fortunately, in our formalism, one can take the local gravitational fields into considerations by giving physical 4-velocities in Eq.~(\ref{29}). It seems obvious that the geometric factors of PTA should contain physical information about the motions of the composed pulsars.

It would be confusing that the values for the magnitude of the ORFs in de-Sitter background is shown to be divergent as $\theta_{\alpha \beta} \rightarrow 0$. It indicates that the auto-correlations of the output of GW detectors should be divergent.  In this sense, difficulties might exist in estimation of the sensitivity of GW detectors.

\smallskip
{\it Acknowledgments. }  
The author thanks Prof.~Qing-Guo Huang for useful discussions and remarks on the detectability.

\bibliography{ref}

\

\appendix
\section{Polarization tensor}

In principle, there is an additional the degree of freedom in polarization plane, which should be averaged \cite{Maggiore:2018sht,Tasinato:2022xyq}. In the present paper, we present a consistent derivation by using the relation $\Lambda_{a   b, c   d} =   e^{\lambda}_{a   b} e_{c   d}^{\lambda}$ in Eq.~(\ref{37}).
Here, we will show that this relation can be reproduced from a non-specific choice of polarization vectors. Firstly, we express the transverse operator $\mathcal{T}_{a   b}$ in the form of
\begin{eqnarray}
  \mathcal{T}_{a   b} & = & (\hat{f}_a \hat{f}_b + e_a
  e_b + \bar{e}_a \bar{e}_b) - \hat{f}_a \hat{f}_b\\
  & = & e_a e_b + \bar{e}_a \bar{e}_b
\end{eqnarray}
where the $e_a$ and $\bar{e}_a$ are the two unit polarization vector with respect to the $\hat{f}$. And the the transverse-traceless tensor reduces to be
  \begin{eqnarray}
    \Lambda_{a   b, c   d} & = & \mathcal{T}_{a   c}
    \mathcal{T}_{b   d} -\mathcal{T}_{a   b} \mathcal{T}_{c
      d} +\mathcal{T}_{a   d} \mathcal{T}_{b   c} \nonumber\\
    & = & (e_a e_c + \bar{e}_a \bar{e}_c)
    (e_b e_d + \bar{e}_b \bar{e}_d) \nonumber  \\ &&- 
    (e_a e_b + \bar{e}_a \bar{e}_b)
    (e_c e_b + \bar{e}_c \bar{e}_b) \nonumber  \\ && +
    (e_a e_d + \bar{e}_a \bar{e}_d)
    (e_b e_c + \bar{e}_b \bar{e}_c)\nonumber\\
    & = & (e_a e_b - \bar{e}_a \bar{e}_b)
    (e_c e_d - \bar{e}_c \bar{e}_d) \nonumber\\ && +
    (e_a \bar{e}_b + \bar{e}_a e_b)
    (e_c \bar{e}_d + \bar{e}_c e_d)
  \end{eqnarray}
By introducing
\begin{eqnarray}
  e^+_{a   b} & \equiv & e_a e_b - \bar{e}_a
  \bar{e}_b\\
  e^{\times}_{a   b} & \equiv & e_a \bar{e}_b + \bar{e}_a
  e_b
\end{eqnarray}
One can reproduce the relation $\Lambda_{a   b, c   d} =
  e^{\lambda}_{a   b} e_{c   d}^{\lambda}$ used in Eq.~(\ref{37}).

\

\section{Perturbed geodesic equations} \label{B}
  The geodesic equations are given by
  \begin{eqnarray}
    \nabla_p p^{\mu} & = & 0~,
  \end{eqnarray}
  Based on the expansion $p^{\mu} \rightarrow  {p}^{\mu} + \delta p^{\mu}$, and ${g}_{\mu \nu}\rightarrow g_{\mu \nu} +  \delta g_{\mu \nu}$,
  we have
  \begin{eqnarray}
    \hspace*{\fill} p^{\mu} \nabla_{\mu} p^{\nu} & = &
    ( {p}^{\mu} + \delta p^{\mu}) (\partial_{\mu} ( {p}^{\nu} + \delta
    p^{\nu}) + \Gamma_{\mu \lambda}^{\nu} ( {p}^{\lambda} + \delta
    p^{\lambda}))\nonumber\\
    & = &  {p}^{\mu}  {\nabla}_{\mu}  {p}^{\nu} + \delta p^{\mu}
      {\nabla}_{\mu}  {p}^{\nu} +  {p}^{\mu}  {\nabla}_{\mu} \delta
    p^{\nu} \nonumber\\ && + g^{\nu \rho} \left( 2 \nabla_{\mu}  \delta g_{\lambda \rho} - \frac{1}{2}
    \nabla_{\rho}  \delta g_{\mu \lambda} \right)  {p}^{\mu}  {p}^{\lambda}~, \label{B2}
  \end{eqnarray}
  in which, we have used the expansion of Christoffel symbols, 
  \begin{eqnarray}
    \Gamma^{\nu}_{\mu \lambda} & = & \frac{1}{2} g^{\nu \rho}
    (\partial_{\mu} g_{\lambda \rho} + \partial_{\lambda} g_{\mu
      \rho} - \partial_{\rho} g_{\mu \lambda})\nonumber\\
    & = & \Gamma^{\nu}_{\mu \lambda} + \frac{1}{2} h^{\nu \rho} (\partial_{\mu}
    g_{\lambda \rho} + \partial_{\lambda} g_{\mu \rho} - \partial_{\rho} g_{\mu
        \lambda}) \nonumber\\
    & & + \frac{1}{2} g^{\nu \rho} (\nabla_{\mu}  \delta g_{\lambda \rho} + \Gamma_{\mu
      \lambda}^{\kappa}  \delta g_{\kappa \rho} + \Gamma^{\kappa}_{\mu \rho}  \delta g_{\lambda
        \kappa} 
    \nonumber \\ & & + \nabla_{\lambda}  \delta g_{\mu \rho} + \Gamma^{\kappa}_{\lambda \mu}
    \delta g_{\kappa \rho} + \Gamma^{\kappa}_{\lambda \rho}  \delta g_{\mu \kappa} 
    \nonumber \\ & & -
    \nabla_{\rho}  \delta g_{\mu \lambda} - \Gamma_{\rho \mu}^{\kappa}  \delta g_{\kappa
        \lambda} - \Gamma^{\kappa}_{\rho \lambda}  \delta g_{\mu \kappa})\nonumber\\
    & = & \Gamma^{\nu}_{\lambda} + \frac{1}{2} g^{\nu \rho} (\nabla_{\mu}
    \delta g_{\lambda \rho} + \nabla_{\lambda}  \delta g_{\mu \rho} - \nabla_{\rho}  \delta g_{\mu
        \lambda})~, \nonumber\\
  \end{eqnarray}
  The $\nabla_\mu$ is covariant derivative with respect to background metric $g_{\mu\nu}$.

  From Eq.~(\ref{B2}), we obtain the first order geodesic equations,
  \begin{eqnarray}
    {p}^{\mu}  {\nabla}_{\mu}  {p}^{\nu} & = & 0~,
  \end{eqnarray}
  and the second order geodesic equations,
  \begin{eqnarray}
    && \delta p^{\mu}  {\nabla}_{\mu}  {p}^{\nu} +  {p}^{\mu}
    {\nabla}_{\mu} \delta p^{\nu}  \nonumber \\ && + g^{\nu \rho} \left( \nabla_{\mu}
    \delta g_{\lambda \rho} - \frac{1}{2} \nabla_{\rho}  \delta g_{\mu \lambda} \right)
    {p}^{\mu}  {p}^{\lambda}  = 0~.
  \end{eqnarray}

\end{document}